# Measurement of breast-tissue x-ray attenuation by spectral mammography: solid lesions

Erik Fredenberg,[1] Fleur Kilburn-Toppin,[2] Paula Willsher,[2] Elin Moa,[1] Mats Danielsson,[4] David R Dance,[5,6] Kenneth C Young[5,6] and Matthew G Wallis[2,3]

[1] Philips Healthcare, Mammography Solutions, Smidesvägen 5, 171 41 Solna, Sweden
[2] Cambridge Breast Unit, Addenbrookes Hospital, Hills Road, Cambridge CB2 0QQ, United Kingdom
[3] NIHR Cambridge Biomedical Research Centre, Addenbrookes Hospital, Hills Road, Cambridge CB2 0QQ, United Kingdom
[4] Department of Physics, Royal Institute of Technology (KTH), AlbaNova University Center, 106 91 Stockholm, Sweden
[5] NCCPM, Royal Surrey County Hospital, Guildford GU2 7XX, United Kingdom
[6] Department of Physics, University of Surrey, Guildford GU2 7XH, United Kingdom

E-mail: erik.fredenberg@philips.com

**Abstract**
Knowledge of x-ray attenuation is essential for developing and evaluating x-ray imaging technologies. For instance, techniques to distinguish between cysts and solid tumours at mammography screening would be highly desirable to reduce recalls, but the development requires knowledge of the x-ray attenuation for cysts and tumours. We have previously measured the attenuation of cyst fluid using photon-counting spectral mammography. Data on x-ray attenuation for solid breast lesions are available in the literature, but cover a relatively wide range, likely caused by natural spread between samples, random measurement errors, and different experimental conditions. In this study, we have adapted the previously developed spectral method to measure the linear attenuation of solid breast lesions. A total of 56 malignant and 5 benign lesions were included in the study. The samples were placed in a holder that allowed for thickness measurement. Spectral (energy-resolved) images of the samples were acquired and the image signal was mapped to equivalent thicknesses of two known reference materials, which can be used to derive the x-ray attenuation as a function of energy. The spread in equivalent material thicknesses was relatively large between samples, which is likely to be caused mainly by natural variation and only to a minor extent by random measurement errors and sample inhomogeneity. No significant difference in attenuation was found between benign and malignant solid lesions, or between different types of malignant lesions. The separation between cyst-fluid and tumour attenuation was, however, significant, which suggests it may be possible to distinguish cystic from solid breast lesions, and the results lay the groundwork for a clinical trial. In addition, the study adds a relatively large sample set to the published data and may contribute to a reduction in the overall uncertainty in the literature.

## 1. Introduction

Basic knowledge of the x-ray attenuation of tissue is essential for the development of new x-ray imaging technologies, as well as the study of the performance of existing technologies. Knowledge of the x-ray attenuation of tissue is particularly important for the development of new applications of unenhanced spectral imaging. Unenhanced spectral imaging is an emerging x-ray imaging technology that measures tissue properties, without injection of a contrast agent, using differences in attenuation as a function of energy. In the field of mammography, unenhanced spectral imaging has been employed to improve the image signal-to-noise ratio (Tapiovaara and Wagner, 1985, Cahn *et al*, 1999), improve lesion visibility (Johns and Yaffe, 1985, Kappadath and Shaw, 2008, Taibi *et al*, 2003, Fredenberg *et al*, 2010b), and to measure breast density (Ding and Molloi, 2012).

Another potential application of unenhanced spectral imaging is to distinguish cysts from tumours and thereby address the relatively low specificity of x-ray mammography screening (Norell *et al*, 2012, Erhard *et al*, 2014). Round lesions are a common mammographic finding, and it can be difficult to distinguish benign cysts from tumours on mammography alone, particularly when the margin is partly obscured. Recall rates for assessment are approximately 5% in Northern Europe and much higher in the USA (Smith-Bindman *et al*, 2003). Not only is this costly for the screening programme (Guerriero *et al*, 2011), but recalls can be stressful for patients (Brett and Austoker, 2001). The evidence that women may be put off returning for subsequent screens following false positive recalls is mixed (Brewer *et al*, 2007, Román *et al*, 2001, Maxwell *et al*, 2013), but this potentially reduces the long-term effectiveness of the screening programme. Improving lesion characterisation at screening would be very desirable to reduce recalls, but the development of spectral x-ray techniques for this purpose has been hindered by a lack of tissue attenuation data.

A few studies have reported measurements of the linear attenuation coefficient of solid breast lesions, mainly with the purpose of comparing to normal fibro-glandular tissue. Pioneering work was done by Johns and Yaffe (1987) who measured the linear attenuation coefficient (energy range 18 to 110 keV) of 6 samples of infiltrating ductal carcinoma using a broad x-ray spectrum and a spectroscopy detector. It was concluded that the attenuation of carcinoma was significantly different from that of fibro-glandular breast tissue at x-ray energies below 40 keV. Carroll *et al* (1994) measured the linear attenuation coefficient (energy range 14 to 18 keV) of 12 samples of malignant breast tumors using monochromatic synchrotron radiation. The study reported a difference in mean values between normal and cancerous tissue, but there was overlap of the distributions. Chen *et al* (2010) measured the linear attenuation coefficient (energy range 15 to 26.5 keV) of ≤14 samples (not all samples were imaged at all energies) of mainly ductal carcinoma using computed tomography with monochromatic synchrotron radiation. The study did not find a significant difference between cancerous and fibro-glandular tissue. Tomal *et al* (2010) measured the linear attenuation coefficient (energy range 8 to 30 keV) of 18 invasive ductal carcinomas and 6 fibroadenomas using an x-ray tube and a silicon monochromator. A significant difference in attenuation between normal and cancerous tissue was found below 28 keV.

We have previously developed a method to measure the energy-dependent x-ray attenuation of tissue samples using spectral imaging (Fredenberg *et al*, 2013). The method is based on mapping of the attenuation to equivalent thicknesses of two reference materials. An advanced prototype clinical spectral-imaging mammography system was used together with the method to measure the attenuation of cyst fluid, which marked the first step in our efforts to evaluate the feasibility of using unenhanced spectral imaging to distinguish between cysts and tumours in screening images. The method was validated by comparing measurements with calculations of attenuation from the measured elemental composition of cyst fluid and the known elemental composition of water.

The purpose of the current study is to measure the attenuation of formalin-fixed solid breast lesion specimens with the same methodology as for cyst fluid, to establish whether the attenuation of solid lesions is different enough from the attenuation of cystic lesions so that the two lesion types could be distinguished from one another by spectral imaging. Such a comparison is difficult to do with published data based on a different measurement setup because the potential bias from the different setups might obscure the small attenuation differences between the tissue types. However, data on tissue attenuation may also have general value for the scientific community as the sources of such data are sparse.

## 2. Materials and Methods

*2.1. Spectral mammography system*
The Philips MicroDose SI mammography system comprises a tungsten-target x-ray tube with aluminium filtration, a pre-collimator, and an image receptor, which is scanned across the object (Figure 1, left). The image receptor consists of photon-counting silicon strip detectors with corresponding slits in the pre-collimator (Figure 1, right). This multi-slit geometry rejects virtually all scattered radiation (Åslund *et al*, 2006). 32 kV acceleration voltage was used for all measurements presented in this study.

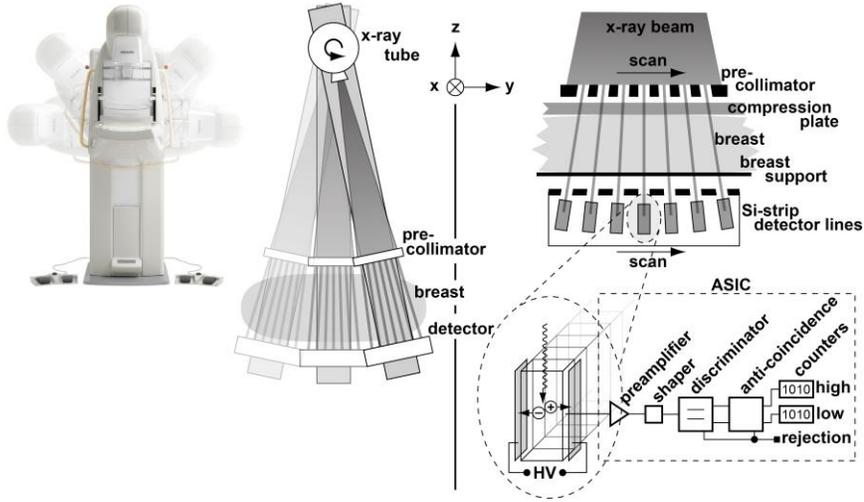

**Figure 1: Left:** Photograph and schematic of the Philips MicroDose Mammography system. **Right:** The spectral image receptor and electronics.

Photons that interact in the detector are converted to pulses with amplitude proportional to the photon energy. A low-energy threshold rejects virtually all pulses below a few keV, which are considered to be generated by noise. An additional high-energy threshold sorts the detected pulses into two bins according to energy. The threshold was set close to 20 keV, which yields approximately equal number of counts in each bin for a typical transmitted spectrum. The energy resolution at this energy level is approximately 5 keV full width at half maximum (Fredenberg *et al*, 2010a).

*2.2. Background of spectral imaging*

For most natural body constituents at mammographic x-ray energies, it is fair to ignore absorption edges. X-ray attenuation is then made up of only two interaction effects, namely photoelectric absorption and scattering processes (Alvarez and Macovski, 1976, Lehmann *et al*, 1981, Johns and Yaffe, 1985). Accordingly, in the mammographic energy range, a linear combination of any two materials of different and low atomic number can approximately simulate the energy-dependent attenuation of a third material of a given thickness,

$$t_{\text{sample}}\mu_{\text{sample}}(E) = t_1\mu_1(E) + t_2\mu_2(E). \qquad (1)$$

We call these materials reference materials, and if this relationship is assumed to hold exactly, then the associated normalized reference thicknesses $[t_1, t_2]/t_{\text{sample}}$ are unique descriptors of the energy dependent sample attenuation ($\mu_{\text{sample}}$) given the known attenuations of the reference materials ($\mu_1$ and $\mu_2$). Further, and with the same assumption, the detected signal ($I$) in a photon-counting x-ray detector would be identical for a tissue sample and for the equivalent combination of reference materials, regardless of incident energy spectrum ($\Phi(E)$) or detector response ($\Gamma(E)$),

$$\begin{aligned} I_{\text{sample}} &= I_0 \int \exp[-t_{\text{sample}}\mu_{\text{sample}}(E)]\Phi(E)\Gamma(E)\mathrm{d}E \\ &= I_0 \int \exp[-t_1\mu_1(E) - t_2\mu_2(E)]\Phi(E)\Gamma(E)\mathrm{d}E = I_{\text{reference}}. \end{aligned} \qquad (2)$$

Hence, measurements of attenuation at two different energies (or for two different energy spectra) yield a non-linear system of equations, which, for known $t_{\text{sample}}$, can be solved for $t_1$ and $t_2$. Measurements at more than two energies yield an over-determined system of equations under the assumption of only two independent interaction processes, and would, in principle, be redundant. Equations (1) and (2) assume that scattering processes can be treated as absorption, which is true only for x-ray detector geometries with efficient scatter rejection, such as multi-slit (Åslund *et al*, 2006).

In this work, aluminium (Al) and polymethyl methacrylate (PMMA) have been used as the two reference materials. In common with previous studies (Lehmann *et al*, 1981, Johns and Yaffe, 1987), it is useful to express the equivalent Al and PMMA thicknesses in terms of the corresponding Al-PMMA vector with magnitude and angle given by

$$r = \sqrt{t_{Al}^2 + t_{PMMA}^2} \quad \text{and} \quad \theta = \tan^{-1}\left(\frac{t_{Al}}{t_{PMMA}}\right). \tag{3}$$

The magnitude $r$ is directly proportional to the thickness and the density (specific weight) of the sample, whereas the angle $\theta$ is related to the attenuation energy dependence and the (effective) atomic number of the material, and is independent of sample thickness.

It may be useful (in this study for comparing to previous measurements) to convert a linear attenuation coefficient to equivalent material thicknesses. In principle, the linear attenuation coefficient could then be related to reference thicknesses $t_1$ and $t_2$ for Al and PMMA by calculating Eq. (1) at two arbitrary energies and solving the resulting linear system of equations. However, the limited validity of the assumptions underpinning Eq. (1) may lead to a slight dependence on the energies chosen. Instead, we used a linear least-squares fit over the x-ray spectrum energy bins $(E_1 \ldots E_N)$, weighted with the expected spectrum after the sample $(\phi(E))$ and the detector response $(\Gamma(E))$. This procedure corresponds to the following minimization with respect to $t_1$ and $t_2$:

$$\min_{t_1,t_2} \sum_{n=1}^{N} \phi(E_n)\Gamma(E_n)\left[t_1\mu_1(E_n) + t_2\mu_2(E_n) - t_{sample}\mu_{sample}(E_n)\right]^2. \tag{4}$$

Weighting with the spectrum is a second-order correction, which was found to influence the calculated reference thicknesses in the order of one percent compared to equal weighting at all energies, and a relatively simple model without experimental validation was therefore deemed accurate enough to calculate $\phi$ and $\Gamma$. For $\phi$, we used a generic tungsten spectrum model (Boone *et al*, 1997), simply attenuated by 0.5 mm Al and the appropriate amount of PMMA from the sample holder. In $\Gamma$, all detector effects were ignored except for quantum efficiency, which was calculated by assuming a 3-mm-thick silicon detector.

*2.3. Spectral attenuation measurements*

As mentioned above, we have previously developed a method to measure x-ray attenuation of tissue samples by solving Eq. (2) for $t_1$ and $t_2$ in spectral images acquired with the Philips MicroDose SI system (Fredenberg *et al*, 2013). The method was applied in this work to solid tissue samples in addition to the liquid cyst-fluid samples investigated previously.

In total, 84 formalin-fixed tissue samples of solid benign and malignant breast lesions were included in the study. Ethical approval was obtained to use samples from women from whom generic consent had been obtained prior to surgery. The generic consent allowed tissue to be used for ethically approved research, including genetic profiling. Immediately post-surgery the tissue was sliced to facilitate even fixation with formalin. For inclusion in this study, sliced samples were obtained from pathology the morning after surgery. Only samples that had the solid breast lesion clearly visible on both surfaces of the sample slice were used to avoid interference from normal breast tissue.

The samples were placed in a hollow PMMA cylinder with a threaded lid, henceforth referred to as the sample holder. By twisting the lid, the sample was gently compressed and the air gap between sample and lid was removed. A thin non-moving PMMA plate between the sample and the threaded lid allowed compression without twisting or otherwise damaging the sample. Figure 2 illustrates the measurement setup.

The sample thickness ($t_{sample}$) was measured using two methods: (1) A protractor on the threaded lid allowed thickness measurement with a resolution of approximately 10 µm by calibrating the protractor on a series of gauge blocks, and (2) the height of the cylinder was measured with a calliper, also with a resolution of 10 µm, from which the known bottom and lid thicknesses were subtracted. The means of the two thickness measurements were used in the calculation to produce normalized values of $t_{Al}$ and $t_{PMMA}$.

The samples were imaged with the mammography stand in horizontal mode. An Al and PMMA step wedge was positioned adjacent to the sample and was present in each image to provide a reference grid of thickness/material combinations for comparison to the sample. X-ray attenuation was measured by mapping the high- and low energy counts obtained from a region-of-interest (ROI) located on the sample ($I_{sample}$) against those obtained from ROIs on the step wedge ($I_{reference}(t_{Al}, t_{PMMA})$). Linear Delaunay triangulation in the log domain was used to find intermediate reference values. The range of the step wedge was adapted to the sample thickness by adding thin PMMA plates with an Al sheet on top, referred to as step-wedge plates. For each sample the appropriate number of step wedge plates was added so that $I_{sample}$ fell within $I_{reference}$.

ROI selection was done by a medical physicist and verified / adjusted by an experienced radiologist. Care was taken to avoid areas with visible microcalcifications. Each manually selected ROI was automatically divided into four equal-sized sub ROIs in order to estimate inhomogeneities within a sample, and four images with identical ROI selection were acquired of each sample to estimate random fluctuations between measurements. The variation between the sub ROIs in each image, averaged over all images of the sample, is referred to as the intra-image variability. The variation between sub ROIs with equal position but in different images, averaged over all sub ROI positions, is referred to as the inter-image variability. Data from all ROI locations and images (four-by-four individual readings) were added in order to estimate the expectation value for each sample. The spread between the expectation values for different samples is referred to as the total variability. Image variability measures are overlaid with quantum noise, which was estimated from knowledge of the photon counts per pixel in the high- and low-energy images and error propagation through the interpolation process. Two-sample *t*-tests and chi-square variance tests were used to test the differences between the means and the variances of the different measurements. The results were compared to the previously obtained measurements of 50 samples of cyst-fluid and 50 samples of water (Fredenberg *et al*, 2013).

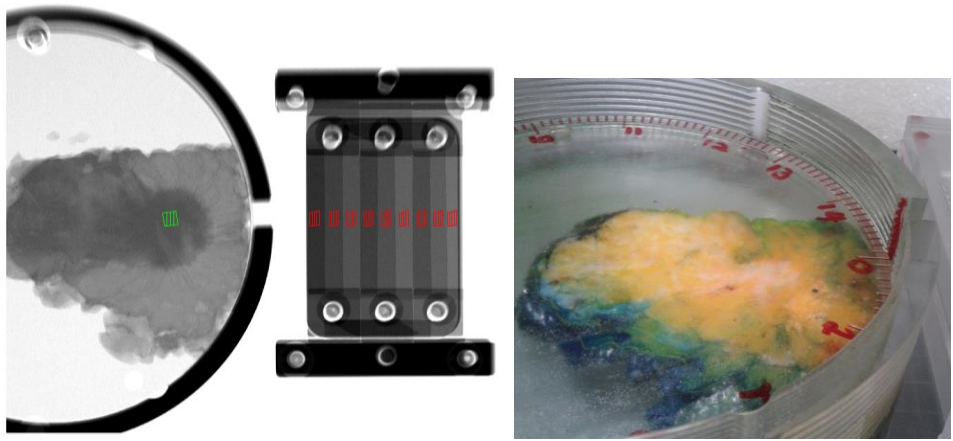

**Figure 2:** X-ray image (left) and photograph (right) of the sample holder and step wedge with a tumour sample in place (grade I invasive ductal carcinoma). Sample ROIs that were used for the evaluation are indicated in green and step-wedge ROIs in red. There were four independent measurement points on each sample, which are illustrated by gridded ROIs.

*2.4. Systematic and random measurement errors*

The thicknesses of the PMMA step wedge and sample holder components were measured with a micrometre screw, and the PMMA density was determined by weighing the step wedge and pieces of sample holder material with high-precision scales and measuring their size with the micrometre screw. The thicknesses of the PMMA mounting plates were measured with a calliper in a grid of positions at the step wedge and sample locations so that variations in the thickness could be taken into account. The PMMA densities of the mounting plates were not measured because the difference in thickness between sample and step wedge locations is small so any reasonable deviation from a standard value would have minimal impact.

Random errors in the PMMA thicknesses were estimated as the standard deviation of measurements at several locations, or as the resolution of the measurement device, whichever was largest. Thickness variations with position were considered random (varying from sample to sample) because the ROI position varied between measurements. A systematic error (identical for all samples) from the measurement device corresponding to the resolution of the device was assumed for each measurement. Systematic PMMA density errors of the sample holder and step wedge were found by propagating weighing and volume errors through the density calculation. The volume and weighing errors were estimated by repeated measurements. The density of the step wedge plates was not measured and a systematic error was estimated as the approximate variation of densities in off-the-shelf PMMA from different manufacturers.

The Al steps in the step wedge were constructed from layers of Al foil (purity 99.999%). The thicknesses of these were determined by measuring the area and weight of an approximately 100×100 mm$^2$ Al sheet, assuming the elemental density, and solving for the thickness. A systematic thickness error was estimated by propagating the weighing and area errors through the calculation. The weighing error were estimated by repeated measurements and the area error was estimated by the resolution of the measurement device (a ruler). The Al sheets on the step-wedge plates were made of the same type of foil, but the thickness was instead measured with a micrometre screw and a systematic error was estimated as the resolution of the device. Random errors in Al thickness for the step wedge and step-wedge plates were estimated as the standard deviation of measurements with a micrometre screw at several locations on the Al sheets.

As the sample thickness measurement was repeated for each individual sample, i.e. the sample thickness was not constant, we can expect a random uncertainty in thickness, which shows up as a scaling of $r$ (cf. Eq. (3)). These random variations were estimated as the standard deviation of the difference between the two thickness measurements (protractor and calliper) for all samples. Note, however, that this error estimation does not capture potential air gaps between the lid and the sample (the samples were not re-compressed between the two measurements), but we assume this source of error to be negligible. The resolution of the protractor and calliper was included as a systematic error.

## 3. Results

Of the original 84 samples, 23 were excluded from the study: 6 cases were DCIS with no mass lesion discernible on the images, and 17 cases were excluded for technical reasons, mainly because the pathology lab detected that tumour was not histologically present at both cut surfaces despite initial appearances of the fixed specimen block. Of the remaining 61 biopsy proven solid fixed tissue specimens, 37 were invasive ductal carcinoma (3 grade I, 19 grade II, 15 grade III), 14 lobular carcinoma, 5 special type (2 papillary carcinoma, 2 mucinous, and 1 metaplastic), and 5 benign (4 fibroadenomas and 1 phyllodes tumour). Due to the logistical constraints of obtaining tissue from, and returning it to, the pathology laboratory cut up room, without disruption or delaying processing, final sample eligibility could only be verified after imaging had been performed, leading to the relatively large number of exclusions.

The mean sample thickness was 6.4 mm, and the mean total sample ROI size (sum of four sub ROIs) was 54 mm$^2$. The systematic errors caused by uncertainties in thickness and density measurements were estimated according to Sec. 2.4 for an average sample to 0.7% in the Al and PMMA equivalent thicknesses. The random errors caused by variations in thickness over the measurement area and variation in the sample thickness measurements were estimated for an average sample to be 0.9% and 0.5% in the Al and PMMA equivalent thicknesses, respectively.

Figure 3 shows measurement results for the 61 solid (5 benign, 56 malignant) samples that were included in this study, as well as for 50 cyst fluid samples and 50 water samples from Fredenberg *et al* (2013), all expressed in terms of the equivalent Al and PMMA thicknesses and normalized to a tissue sample thickness of 10 mm and a PMMA density of 1.19 g/cm$^3$. The left panel of Figure 3 shows an overview of the Al-PMMA vectors with the total variability (one standard deviation) indicated as error bars at the end of each vector. The angle ($\theta$) and magnitude ($r$) are illustrated on the water vector. Note that the Al and PMMA axes in Figure 3 are transposed compared to some previous studies (Johns and Yaffe, 1985, 1987), but we have kept the definition of $\theta$ (denoted $\Phi$ by Johns and Yaffe) to facilitate comparison, which is why the angle runs from the ordinate rather than from the abscissa in Figure 3. The (effective) atomic number of the sample material determines $\theta$, and the angles for $Z = 7$ and $Z = 8$ are indicated in Figure 3 for illustration.

The right panel of Figure 3 shows a close-up of the measurements. For the solid lesions, individual measurement points are shown shaded in grey with error bars for the inter-image variability and the 5 benign lesions are circled. For the previously published cyst fluid and water measurements, the perimeters (convex hulls) of the set of individual measurement points are shown. The mean values and total variabilities of each sample type are shown as bold error bars. The smallest $\theta$ of the cyst fluid samples is indicated with a red dotted line and the most extreme outlier of the solid lesions toward the cyst fluid distribution is indicated with an arrow. All mean values and variability measures are listed in Table 1.

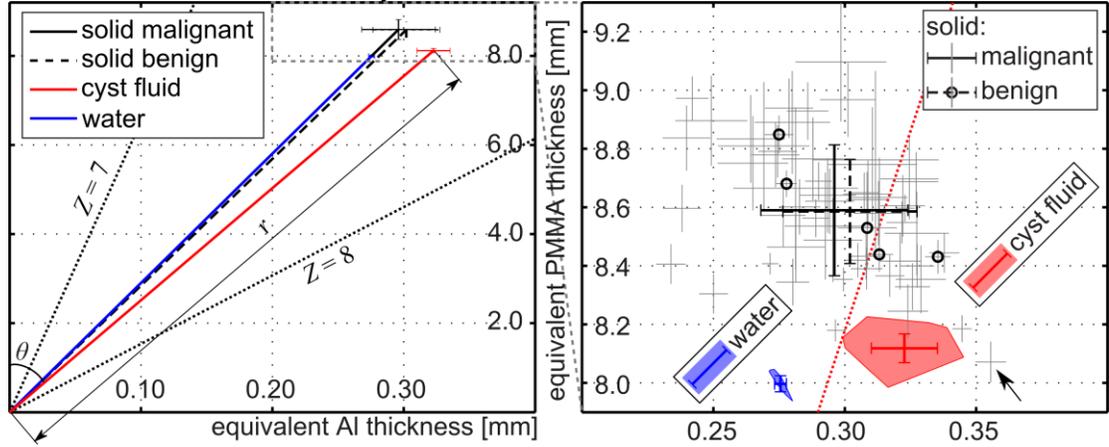

**Figure 3:** Equivalent PMMA and Al thickness normalized to a 10 mm sample, measured on 61 samples of solid lesions (benign and malignant) and compared to previously published data on cyst fluid and water. **Left:** Overview of the Al-PMMA vectors with the angle $\theta$ and magnitude $r$ indicated on the water vector. The total variabilities (one standard deviation) are shown as error bars at the end of each vector. Dotted lines indicate $\theta$ for atomic numbers $Z = 7$ and $Z = 8$. **Right:** Close-up of the measurement points. Each individual measurement is shown for the solid lesions, and the error bars show one standard deviation of the inter-image variability. The 5 benign lesions are circled. For cyst fluid and water, the perimeters (convex hulls) of the set of individual measurement points are shown. The error bars for each sample type (bold) show the mean and standard deviation of the measured values (i.e. the total variability).

**Table 1:** Equivalent PMMA and Al thicknesses measured on 61 samples of solid breast lesions normalized to 10 mm sample thickness (bold) and compared to previously published data on 50 samples of cyst fluid and water. One ROI was manually placed on each sample and automatically divided into four sub ROIs. Four images with identical ROI positioning were acquired of each sample. All variabilities are quantified using the standard deviation in percentage of the mean value. Total variability is the sample-to-sample variability (the variability between the averages of all readings of each sample), which is quantified with the minimum and maximum values of the distribution in addition to the standard deviation (SD). The expected quantum noise contribution to the total variability (total quantum) was calculated from all ROIs and all images of each sample. Inter-image variability is the image-to-image variability of each sample (the average variability between equal sub ROIs in different images), and intra-image variability is the sub ROI variability (the average variability between the sub ROIs in each image). The expected quantum noise (image quantum) is the same for both of these variability measures. Additionally, the angle ($\theta$) and magnitude ($r$) of the Al-PMMA vectors are given and marked in grey to indicate that these measures are not independent of the PMMA and Al thicknesses.

| Measure | Sample type | Mean | Variability total variability | | | total quantum | inter-image | intra-image | image quantum |
|---|---|---|---|---|---|---|---|---|---|
| | | | SD | Min | max | | | | |
| PMMA [mm] | solid malignant: | 8.59 | 2.6% | 8.07 | 9.10 | 0.59% | 2.6% | 2.3% | 2.4% |
| | solid benign: | 8.59 | 2.1% | 8.43 | 8.85 | 0.41% | 1.7% | 1.8% | 1.7% |
| | cyst: | 8.12 | 0.61% | 7.98 | 8.23 | 0.34% | 1.2% | 1.2% | 1.4% |
| | water: | 8.00 | 0.34% | 7.94 | 8.05 | 0.31% | 1.2% | 1.1% | 1.2% |
| Al [mm] | solid malignant: | 0.296 | 9.5% | 0.234 | 0.356 | 1.5% | 6.2% | 5.9% | 6.1% |
| | solid benign: | 0.302 | 8.4% | 0.275 | 0.335 | 1.0% | 4.1% | 3.8% | 4.1% |
| | cyst: | 0.323 | 3.9% | 0.299 | 0.345 | 0.75% | 2.5% | 2.5% | 3.0% |
| | water: | 0.276 | 0.82% | 0.271 | 0.280 | 0.80% | 2.9% | 2.7% | 3.1% |
| $\theta$ [mrad] | solid malignant: | 34.5 | 9.8% | 27.0 | 44.0 | 1.6% | 6.7% | 6.3% | 6.5% |
| | solid benign: | 35.2 | 8.7% | 31.1 | 39.8 | 1.1% | 4.5% | 4.3% | 4.5% |
| | cyst: | 39.8 | 4.0% | 36.7 | 42.7 | 0.82% | 2.7% | 2.8% | 3.3% |
| | water: | 34.5 | 0.89% | 33.7 | 35.3 | 0.86% | 3.1% | 3.0% | 3.4% |
| $r$ [mm] | solid malignant: | 8.59 | 2.6% | 8.08 | 9.10 | 0.59% | 2.6% | 2.3% | 2.4% |
| | solid benign: | 8.59 | 2.1% | 8.44 | 8.85 | 0.41% | 1.7% | 1.8% | 1.7% |
| | cyst: | 8.12 | 0.61% | 7.99 | 8.23 | 0.34% | 1.2% | 1.2% | 1.4% |
| | water: | 8.00 | 0.34% | 7.94 | 8.05 | 0.31% | 1.2% | 1.1% | 1.2% |

There was no significant difference in attenuation between any of the malignant lesion types (P > 0.5 determined by two-sample *t*-tests on the PMMA and Al equivalent thicknesses), and in common with Tomal *et al* (2010) we have therefore bundled the data for the malignant lesions in the following. There was no significant difference in attenuation between benign and malignant solid breast lesions (P > 0.5 determined by two-sample *t*-tests on the PMMA and Al equivalent thicknesses), which is also manifested by the almost complete overlap by the mean values for benign solids and malignant solid lesions in Figure 3.

Solid breast lesion attenuation (benign and malignant) was significantly different from cyst-fluid attenuation and significantly different from water attenuation in terms of PMMA and Al equivalent thicknesses (P < 0.001, determined by two-sample *t*-tests). Solid breast lesion attenuation was significantly different from cyst attenuation in terms of angle ($\theta$) and magnitude ($r$) of the Al-PMMA vector and significantly different from water attenuation in terms of $r$ (P < 0.001, determined by two-sample *t*-tests), but there was no significant difference from water attenuation in terms of $\theta$ (P > 0.5, determined by a two-sample *t*-test). The small difference between water and solid lesions in terms of $\theta$ is manifested by the Al-PMMA vector for solid lesions in Figure 3 being obscured by the water vector for most but not all of its length as the $r$ values are different.

The random fluctuations between measurements (inter-image) and the sample variability (intra-image) were not significantly different from the expected quantum noise between ROIs (P > 0.1, determined by two-sample *t*-tests between the standard deviation of the measured ROI-to-ROI variation for each sample and the expected standard deviation of the quantum noise), which indicates that quantum noise is the dominating noise source and that the sample homogeneity and system stability were good.

The total variability of the solid samples was, however, significantly larger than the expected quantum noise (P < 0.001, determined by chi-square variance tests on the PMMA and Al equivalent thicknesses). The total variability of the solid malignant samples was 3 (Al) and 4 (PMMA) times larger than that of the cyst fluid samples, and 16 (Al) and 7 (PMMA) times larger than for the water samples.

The linear attenuation coefficients of solid breast lesions, calculated from the measured PMMA and Al equivalent thicknesses at a range of x-ray energies relevant to mammography, are shown in Table 2 and plotted in Figure 4. (Note that Figure 4 does not include the benign lesions as the difference to malignant lesions was too small to be visualized in the plot.) Previously published linear attenuation coefficients of cyst fluid and water, calculated in an equivalent manner (Fredenberg *et al*, 2013), are included in Table 2 and Figure 4 for reference. Figure 4 also plots the linear attenuation coefficients of solid lesions and cyst fluid normalized to the linear attenuation coefficient of water in order to better visualize the difference in attenuation between the materials.

**Table 2:** Linear attenuation coefficients of solid breast lesions (benign and malignant) calculated from the measured PMMA and Al equivalent thicknesses (bold), and compared to previously published data on cyst fluid and water. The total variability (one standard deviation) is given in parenthesis.

| photon energy [keV]: | 15 | 20 | 25 | 30 | 35 | 40 |
|---|---|---|---|---|---|---|
| **malignant solid linear attenuation [cm$^{-1}$]:** | **1.76** (0.054) | **0.859** (0.024) | **0.541** (0.013) | **0.400** (0.009) | **0.327** (0.007) | **0.286** (0.006) |
| **benign solid linear attenuation [cm$^{-1}$]:** | **1.78** (0.035) | **0.864** (0.014) | **0.544** (0.006) | **0.402** (0.003) | **0.328** (0.002) | **0.286** (0.002) |
| cyst fluid linear attenuation [cm$^{-1}$]: | 1.76 (0.029) | 0.852 (0.013) | 0.533 (0.007) | 0.391 (0.004) | 0.318 (0.003) | 0.277 (0.003) |
| water linear attenuation [cm$^{-1}$]: | 1.64 (0.002) | 0.800 (0.001) | 0.504 (<0.001) | 0.372 (<0.001) | 0.305 (<0.001) | 0.266 (<0.001) |

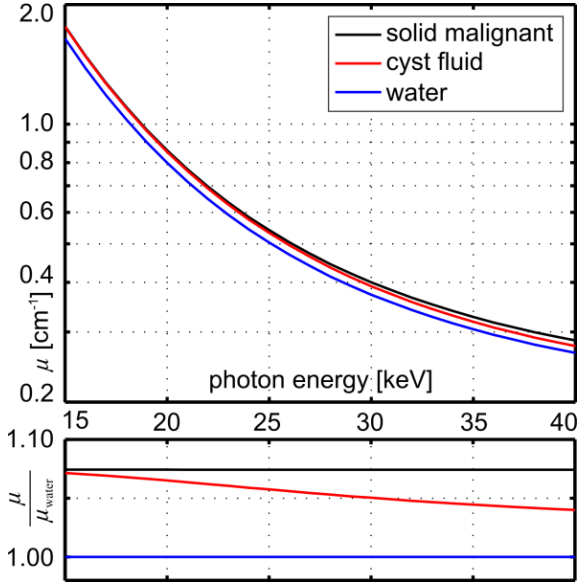

**Figure 4, Top:** Linear attenuation coefficients of malignant solid breast lesions calculated from the measured PMMA and Al equivalent thicknesses, and compared to previously published data on cyst fluid and water. **Bottom:** The linear attenuation coefficients of malignant solid lesions and cyst fluid normalized to the linear attenuation coefficient of water.

## 4. Discussion

*4.1. Measurement results*

*4.1.1. Differentiation between tissues with spectral imaging*

The challenge of distinguishing between different material types in regular x-ray imaging is that it is not possible to differentiate between a thin piece of highly attenuating material and a correspondingly thicker piece of lower attenuating material. Spectral imaging circumvents this problem by considering not only the actual attenuation, but also the energy dependence of the attenuation. It may therefore be possible to distinguish between cyst fluid and solid lesions despite the close-to overlapping linear attenuation curves in the upper panel of Figure 4 (less than 2% difference on average in the 15-40 keV energy interval) because the shape of the curves are different (manifested as different slopes in the lower panel of Figure 4). Water and solid lesions, on the other hand, may be much more difficult to differentiate between even in spectral imaging, despite the 7% difference in attenuation, because the energy dependence of the linear attenuation is close-to identical (parallel flat curves for solid and water attenuation in the lower panel of
Figure 4).

The differences of water and cyst fluid attenuation relative that of solid lesions are further illustrated by the magnitude ($r$) and angle ($\theta$) of the Al-PMMA vector. As defined toward the end of Sec. 2.2, $r$ is directly proportional to the thickness of the sample and therefore cannot be measured unless the thickness of the sample is known. Such an independent thickness measurement was used in the present pre-clinical study, but may not be available in clinical applications of spectral imaging. Contrary to $r$, $\theta$ is independent of the sample thickness and can therefore be measured with spectral imaging (but not with conventional x-ray imaging) without any a-priori information. It was found in this study that $r$ of water and cyst fluid are both significantly lower than that of solid lesions, whereas $\theta$ is equal for water and solid lesions but significantly higher for cyst fluid compared to solid lesions. It appears that solid lesions have an effective atomic number similar to that of water, but with higher density. The effective atomic number of cyst fluid seems slightly higher than that of solid lesions and water.

Despite the large spread, no solid samples fall within the shaded region of the cyst distribution (Figure 2). Nevertheless, 16 malignant (29% of the total number) and 2 benign (40% of the total number) samples overlap with the cyst distribution in terms of $\theta$ (Figure 2, area below the dotted line), and may therefore be challenging to distinguish from cyst fluid without providing thickness information. One (1) solid sample (Figure 2, arrow) falls "on the other side" of the cyst distribution with a higher $\theta$ than any of the cyst fluid cases.

*4.1.2.  Total variability*

Compared to the previous measurements on cyst fluid, the spread between the different samples of solid breast lesions was large. The random measurement errors (Sec. 2.4) were substantially larger for the measurement of solid lesion attenuation in this study compared to the previous measurements on cyst fluid and water, which is mainly due to the repeated thickness measurements and to the thinner samples. The random fluctuations within each sample (intra-image) and between repetitive measurements (inter-image) were not significantly different from the expected quantum noise, which indicates that quantum noise was the dominating noise source within a measurement, and that the sample homogeneity and system stability were good. The expected quantum noise was almost a factor of two higher for the solid lesions compared to the cyst fluid measurements, again because of the thinner samples.

Put together, the random measurement errors and quantum noise constitute approximately 30% and 20% of the total variability in PMMA and Al equivalent thicknesses, respectively. In contrast, a similar estimation for the water measurements, for which sample variations can be expected to be at a minimum, yields contributions by random measurement errors and quantum noise of more than 90%. Despite good sample homogeneity within each ROI and exclusion of samples without clear tumour surfaces on both sides, it is reasonable to expect that some part of the variability comes from contamination of other tissue types in the samples (mainly adipose and fibro-glandular tissue). Care was, however, taken to minimize this effect (samples without clear tumour surfaces on both sides were excluded, and the ROIs were selected in homogeneous areas of the sample) and sample homogeneity within each ROI was found to be good. We therefore expect the major part (more than 50%) of the total variability to be caused by the natural variation of tumour tissue.

*4.2. Comparison to other studies*

*4.2.1.  Malignant lesions*

Figure 5 shows a comparison between the data presented in this study and most of the attenuation data on breast tumors available in the literature (Johns and Yaffe, 1987, Chen *et al*, 2010, Tomal *et al*, 2010). The data from Carroll *et al* (1994) were not tabulated in the paper and are therefore not included in Figure 5. The equivalent Al and PMMA thicknesses for the published data were found by fitting to the linear attenuation coefficients according to Eq. (4). For all three attenuation data sets, the average residuals of the fitted curve compared to the original one were within 0.005% in the considered energy range. The values calculated from each published study are shown as dark blue open markers, and the mean and standard deviation of all previously published studies are shown as a dark blue error bar. The data from the present study are shown as the convex hulls, and solid and dotted error bars that indicate the mean values and the total variabilities of the malignant and benign lesion distributions. The mean and standard deviation of all published studies, including the present study, are shown as a bold green error bar. The interpretation of the cyan markers in Figure 5 is explained in the following.

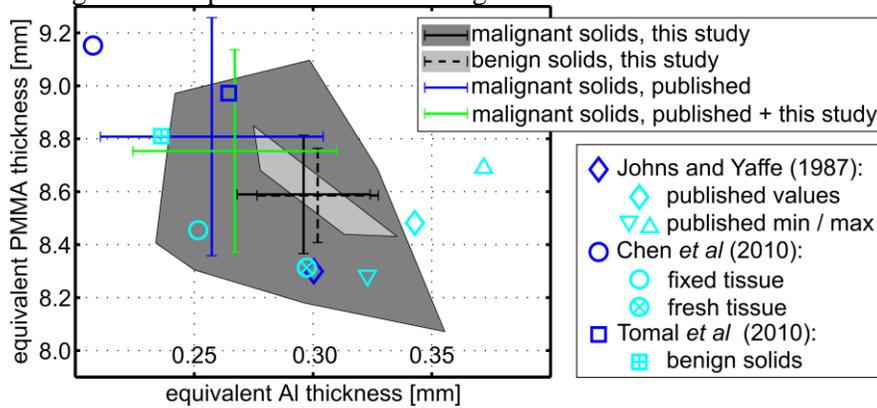

**Figure 5:** The data presented in this study compared to the Al and PMMA equivalent thicknesses calculated from published data on breast tumors (Johns and Yaffe, 1987, Chen *et al*, 2010, Tomal *et al*, 2010). Note that the scaling of the axes are identical to those of Figure 3 (right) and the plots are therefore directly comparable. Data from the published studies are shown as markers with the main results for solid malignant lesions in dark blue, and the data from the present study are shown as the convex hulls of the measurement points. Bold error bars indicate the mean values and variabilities (one standard deviation). Additional data are included as cyan markers: (1) An independent calculation of Al and PMMA thicknesses with maximum and minimum values, (2) Al and PMMA thicknesses for benign lesions, and (3) a comparison between fresh and fixed specimens.

Johns and Yaffe (1987) published the equivalent Al and PMMA thicknesses fitted over the range 18 to 110 keV and these data (renormalized to a thickness of 10 mm and Al and PMMA densities of 2.699 g/cm$^3$ and 1.19 g/cm$^3$) are included in Figure 5 for comparison (cyan diamond). There are clear differences in mean values compared to our fit in the mammography energy range (dark blue diamond). Equation (1) is an approximation and results differ somewhat when fitting over different energy ranges, but that cannot explain the discrepancy; recalculating the linear attenuation from the fit yields an average residual of -0.7% compared to the published linear attenuation over the entire energy interval, whereas a recalculation from the published Al and PMMA thicknesses yields an average residual of 4.0%. This is substantially larger than the maximum fitting error reported by Johns and Yaffe (0.2%), and we therefore expect the major part of the discrepancy to come from errors in the linear attenuation values of Al and PMMA. Firstly, the coherent and incoherent scattering cross sections for compounds are not exactly the weighted sums of the cross sections for individual elements, which affects calculation of the PMMA attenuation. Further, it is conceivable that systematic differences between elemental cross section tabulations have contributed to the error; Johns and Yaffe used the tabulations by Plechaty *et al* (1975) whereas we used those of the XCOM database (Berger *et al*, 2010). Errors in elemental cross section tabulations are typically a few percent and hence not negligible (Saloman *et al*, 1988, Hubbell, 1999).

The equivalent material thicknesses for malignant solid lesions presented in the present study (dark grey convex hull and black solid error bar) are comparable to, and reasonably close to, the mean of previously published data (dark blue markers and error bar). The range of the previously published data is, however, relatively wide. One part of the spread may be caused by randomness within each study, including random measurement errors, quantum noise, and natural variation of tumor tissue. Johns and Yaffe (1987) published the range (maximum and minimum values) of their measured Al and PMMA thicknesses, which are plotted in Figure 5 (cyan diamond and triangles) and gives some indication of the contribution from random errors. Our data on malignant solid lesions cover a wider range (dark grey convex hull), but the set is also substantially larger, which generally increases the range because of outliers but does not affect the range in terms of the standard deviation. (Note that the range of the smaller set of benign lesions from this

study – light grey convex hull in Figure 5 – is similar to that of Johns and Yaffe, but the standard deviation for the benign lesions – dotted error bar – is comparable to our results for malignant lesions – solid black error bar.) Chen *et al* (2010) and Tomal *et al* (2010) reported values of the standard deviation for their measured linear attenuation coefficients. Averaging that data in the range 15 to 30 keV yields values of 0.019 cm$^{-1}$ (Chen *et al*) and 0.061 cm$^{-1}$ (Tomal *et al*), which may be compared to our data on the standard deviation in Table 2 and an average value of 0.022 cm$^{-1}$ for the same energy range. We therefore expect the spread between measurements (the total variation) seen in the present study to be of the same order as that of previous studies.

If random fluctuations within each published study would be the sole contributor to the spread between the mean values, and assuming equal random errors in the present study and in the published studies on average, the spread (standard deviation) between the published mean values would be smaller than the total variation reported in this study because the mean values are better estimates of the expectation value then are the individual measurements. The spread between published mean values is, however, approximately a factor of two *larger* than the total variation of the present study, and it is likely that a substantial part is caused by systematic differences between the measurement setups. We estimated our systematic measurement errors to be 0.7%, and it is likely that other studies end up at similar values, but there may also be other systematic differences between the studies, including differences in tissue handling and tumor types, and in our case, the aforementioned uncertainty in the elemental cross section tabulations. One way of circumventing these types of systematic uncertainties for quantitative analyses is to perform all measurements within the same framework. We have therefore measured tumor attenuation with the same method as previously applied to cyst fluid and under conditions that are as similar as possible to the screening environment where potential future applications will be used.

In summary, the spread in the available attenuation data may be caused by 1) a large natural spread between samples, 2) random measurement errors, and 3) different experimental conditions in the different studies that cause systematic differences and errors. Uncertainty number 1) and 2) may be reduced by adding *samples* to the available data. The present study contributes a relatively large sample set (56 samples) compared to published studies (~38 samples in total) and may help in that respect. Further, assuming that the systematic errors within each study are random between the studies, simply adding more *studies* to the literature will help reduce uncertainty number 3). Including the present study shifts the average of published data slightly (green error bar in Figure 5), and it is fair to assume that to be a better estimate of the true expectation value.

### 4.2.2. Benign lesions

Tomal *et al* (2010) measured the attenuation of 6 fibroadenomas, and the calculated equivalent Al and PMMA thicknesses are shown in Figure 5 (cyan square and cross). There is a clear difference to the mean value of benign lesions reported in this study (dotted error bar), which is similar to the difference between malignant samples in the two studies (solid black error bar and dark blue square) and is likely caused by systematic differences between the measurements. Further, the data reported by Tomal *et al* suggest slightly lower Al and PMMA thicknesses compared to malignant samples, as opposed to our study, where the Al and PMMA thicknesses were rather higher or equal to the malignant samples. Similar to our study, however, Tomal *et al* reported that the difference was not significant, and these discrepancies may be caused by statistical errors. It should be noted that three samples of benign lesions were included also in the study by Johns and Yaffe (1987), but data for these samples were not tabulated separately.

### 4.2.3. Tissue fixation

In this study, formalin fixed tissue was used as opposed to fresh tissue, which is a potential bias when making direct interpretation of the results to clinical mammography. Chen *et al* (2010) measured the linear attenuation coefficients in the energy range 17 to 23 keV of 6 of their malignant samples before and after formalin fixation. The resulting fitted Al and PMMA equivalent thicknesses are shown in Figure 5 (cyan circles). The mean value of the fixed tissue (cyan open circle) is clearly different from that of the full study (dark blue circle). This difference may be caused by the use of a smaller number of samples and the narrow energy span. More interesting is the difference between fixed (cyan open circle) and fresh (cyan circle and cross) tissue within the subset; fresh tissue exhibits a higher Al thickness and a lower PMMA thickness than fixed tissue, which moves the solid lesions closer to the attenuation of cyst fluid. This is slightly worrying for the application of separating solid lesions from cysts, but it is currently not clear how significant the difference between fresh and fixed tissue is considering in particular the calculation of Al and PMMA thicknesses within the limited energy range.

## 5. Conclusions

A significant separation between cyst fluid and tumour attenuation was found on average, which suggests it may be possible to distinguish cystic from solid breast lesions using screening with spectral mammography, raising the exciting possibility of reducing unnecessary recalls. As our study measured tumour attenuation under conditions that were as similar as possible to the screening environment and with the same spectral method as a previous study on cyst-fluid attenuation, we expect the comparison between cysts and tumours to be valid and applicable to screening applications. These results therefore lay the groundwork for a clinical trial, which is currently under way.

Previously published data on the x-ray attenuation of solid breast tumours cover a relatively wide range, which is likely to be caused by 1) a large natural spread between samples, 2) random measurement errors, and 3) different experimental conditions in the different studies that cause systematic differences and errors. The present study adds a relatively large sample set to the published data and may contribute to reduce the overall uncertainty in the literature.

There was a relatively large spread between the different samples of solid breast lesions. We expect the major part of the spread to be caused by natural variation and only to a minor extent by random measurement errors and sample inhomogeneity. No significant difference in attenuation was found between benign and malignant solid lesions, or between different types of malignant lesions.

Our future work will focus on a clinical trial to further investigate the possibility of distinguishing cysts from tumours in the screening image. In addition, our pre-clinical study will continue in order to clarify the difference in x-ray attenuation between fixed and fresh tissue, and to measure the attenuation of healthy adipose and fibro-glandular breast tissue.

**Acknowledgements**

Special thanks are extended to Dr E Provenzano, Breast Histopathologist, and Mr James Neal, Advanced Practitioner in Breast Dissection, both at the Department of Histopathology, Addenbrooke's Hospital and Cambridge NIHR Biomedical Research Centre, for their support in preparing the tissue samples, and to Dr Miriam von Tiedemann, Philips Healthcare, for coordinating parts of the study and for help in methodology development. The authors would also like to thank Dr Björn Cederström and Dr Klaus Erhard, both with Philips Healthcare, for invaluable discussions, as well as Mr Torbjörn Hjärn, Philips Healthcare, and Mr Staffan Karlsson at the Royal Institute of Technology (KTH) for constructing the sample holder. Part of this work was carried out within the OPTIMAM2 project funded by Cancer Research UK (grant number C30682/A17321). The Cambridge Human Research Tissue Bank is supported by the NIHR Cambridge Biomedical Research Centre.